\documentclass[12pt]{article}

\usepackage{amssymb}
\usepackage{amsmath}
\usepackage{amscd}
\usepackage{latexsym}
\usepackage{graphicx}

\usepackage{cite}

\newcommand{\be}{\begin{equation}}
\newcommand{\ee}{\end{equation}}

\newcommand{\bn}{{\bf n}}
\newcommand{\bB}{{\bf B}}
\newcommand{\bS}{{\bf S}}

\newcommand{\bt}{\beta}

\newcommand{\al}{\alpha}

\newcommand{\gm}{\gamma}
\newcommand{\om}{\omega}

\newcommand{\rgl}{\rangle}
\newcommand{\lgl}{\langle}

\begin{document}

\begin{center}

{\Large{\bf Generation of coherent radiation by magnetization \\
reversal in graphene} \\ [5mm]

V.I. Yukalov$^1$, V.K. Henner$^{2,3,4}$, and T.S. Belozerova$^2$  } \\ [3mm]

{\it
$^1$Bogolubov Laboratory of Theoretical Physics, \\
Joint Institute for Nuclear Research, Dubna 141980, Russia \\

$^2$Department of Physics, Perm State University, Perm 614990, Russia \\

$^3$Department of Mathematics, Perm State Technical University, Perm 614990,
Russia 

$^4$Department of Physics, University of Louisville, Louisville,
Kentucky 40292, USA }

\end{center}

\vskip 5cm

\begin{abstract}

Local magnetic moments can be created in graphene by incorporating different defects.
The possibility of regulating dynamics of magnetization in graphene, by employing 
the Purcell effect, is analyzed. The role of the system parameters in magnetization 
reversal is studied. The characteristics of such a reversal can be varied in a wide 
range, which can be used for various applications in spintronics. It is shown that 
fast magnetization reversal generates coherent radiation.  
\end{abstract}

\vskip 5mm


\vskip 2mm


\vskip 2mm

{\it Corresponding author}: V.I. Yukalov, E-mail: yukalov@theor.jinr.ru

\newpage

\section{Introduction}

Finite quantum systems demonstrate a rich variety of properties that can 
be employed for numerous applications. Among such finite systems, it is 
possible to mention quantum dots, nanoclusters, nanomolecules, trapped
atoms, etc. \cite{Birman_1}. A novel class of finite quantum systems 
is presented by graphene. 

When considering electronic properties of graphene, magnetic effects 
are usually disregarded, since they occur on much smaller energy scales 
than other energies 
\cite{Castro_1,Abergel_2,Dresselhaus_38,Goerbig_3,Saito_39,Kotov_4,Katsnelson_5,
Mccann_6,Wehling_7}. 
For instance, the Zeeman energy $g \mu_B B$, in the external field of 
$B = 1$ T, is only $4.3 \times 10^{-5}$ eV ($0.5$ K). This is much smaller 
than other characteristic energies in the tight-binding form of the 
electronic Hamiltonian, the nearest-neighbor hopping energy $2.8$ eV 
($3.25 \times 10^4$ K), the next-nearest-neighbor hopping energy $0.1$ eV
($1.16 \times 10^3$ K, and the electron-electron on-site Coulomb repulsion 
energy that is of order $1 - 10$ eV ($10^4 - 10^5$ K). Direct spin-spin
electron interactions are also small, 
$\mu_B^2/a^3 \approx \rho \mu_B^2/a \sim 10^{-5}$ eV ($0.2$ K), where 
$a \approx 1.42$ \AA\; is carbon-carbon spacing and 
$\rho \approx 3.9 \times 10^{15}$ cm$^{-2}$ is the planar density of 
carbon atoms. The only energy that is smaller than the Zeeman energy
is the energy of spin-orbit interactions, which is of order 
$10^{-6} - 10^{-5}$ eV ($0.01 - 0.1$ K).   

Magnetic effects become more pronounced in the presence of disorder that 
can come about in many different forms, such as adatoms, vacancies, 
admixtures on the top of graphene or in the substrate, and also extended 
defects, such as cracks and edges \cite{Castro_1,Abergel_2,Goerbig_3,Kotov_4,
Katsnelson_5,Mccann_6,Wehling_7,Palacios_46}. 
Adatoms can possess magnetic moments, interacting with electronic spins
as in the Kondo problem \cite{Andrei_8}. Magnetic moments also develop around
vacancies. Spins, localized at such defects, interact with each other, either
ferromagnetically or antiferromagnetically. Magnetization can be induced at
the edges of graphene flakes and quantum dots. A graphene quantum dot of 
radius $R \sim 100$ nm can have $10-20$ edge spins \cite{Kotov_4,Huang_9}. 
A series of quantum dots can form graphene ribbons with disordered edges.

There are numerous works confirming the existence of localized magnetic 
moments at graphene defects, such as zigzag edges 
\cite{Palacios_46,Enoki_10,Jiang_11,Otani_12,Gorjizadeh_13,Magda_14}, 
vacancies 
\cite{Lisenkov_15,Chen_16,Nair_47,Yang_18,Lopez_19,Li_20,Faccio_21}, 
divacancies \cite{Bhandary_48,Jaskolski_40}, adatoms \cite{Power_22,Pike_23,Pike_24},
and transition-metal dimers \cite{Blonsky_49}. Hydrogenated zigzag edges 
\cite{Gorjizadeh_13,Kabir_25} can possess local magnetic moments with spins 
$1/2$, $3/2$, and $5/2$. Such dehydrogenated zigzag-edge groups form dehydrogenated 
nanomolecules exhibiting ferromagnetism at room temperatures 
\cite{Gorjizadeh_13,Qaiumzadeh_26}. Examples of these nanomolecules are: 
C$_{56}$H$_{22}$, C$_{64}$H$_{23}$, C$_{56}$H$_{24}$, C$_{64}$H$_{25}$, 
and C$_{64}$H$_{27}$. Zigzag edges can show magnetism in bilayer graphene \cite{Zhang_27}. 
Organic substrates on graphene can be magnetized \cite{Garnica_28}. Strain can also 
induce magnetism at zigzag edges \cite{Hu_29,Roy_30}. In some cases, defects can 
form magnetic clusters on graphene \cite{Boukhvalov_31}, with spins up to $S = 5/2 $. 
Paramagnetic impurities of isolated Mn$^{2+}$ ions also possess spins $S = 5/2 $ 
\cite{Panich_32}.

Various types of defects in graphene are described in the review articles 
\cite{Terrones_33,Bekyarova_34}. Emergence of defect-induced magnetism in 
graphene materials has been reviewed by Yazyev \cite{Yaziev_35} and by Enoki 
and Ando \cite{Enoki_36}.  

Magnetic properties of graphene nanostructures offer unique opportunities for 
various technological applications related to spintronics, for instance, in
quantum information processing. One of the most important requirements for 
efficient spin manipulation is the possibility to quickly vary the spin 
direction. 

In the present paper, we describe a method allowing for fast magnetization 
reversal in graphene. We accomplish numerical simulations and analyze the 
characteristic features of the reversal process. We show that, by employing 
the Purcell effect \cite{Purcell_37}, that is, coupling the sample to a resonant 
electric circuit, it is possible to efficiently regulate the magnetization 
dynamics. Here it is important to remind that under Purcell effect one often 
understands quantum electrodynamics phenomena exhibiting enhancement of emission 
processes due to a resonant cavity. However, the meaning of the Purcell effect 
is more general, implying the enhancement of relaxation phenomena due to coupling 
with a resonator. Moreover, in his original paper \cite{Purcell_37} Purcell 
considered not a cavity electrodynamics phenomena, but the relaxation of a spin 
system. It is exactly spin dynamics that is the topic of the present paper, and the 
Purcell effect is understood in its original meaning \cite{Purcell_37}. 

We demonstrate that fast magnetization reversal in graphene can generate coherent 
radiation, whose characteristics depend on the properties of magnetic defects.

\section{Model of magnetic graphene}

Defects in graphene interact with each other by means of exchange interactions
\cite{Yaziev_35,Enoki_36} characterized by a Heisenberg Hamiltonian
\be
\label{1}
 \hat H_{def} = -\; \frac{1}{2} \sum_{i\neq j} J_{ij} \left [ \left (
S_i^x S_j^x + S_i^y S_j^y \right ) + \al S_i^z S_j^z \right ] \;  ,
\ee
where $\alpha$ is an anisotropy parameter and $S_j^\nu$ is a $\nu$ component 
of an effective spin of a $j$-th defect, with $j = 1,2,\ldots,N$. We take into 
account the existence of an external magnetic field $B_0$ directed along the 
axis $z$. In order to realize the Purcell effect, the sample is placed inside 
a magnetic coil of an electric circuit, producing a magnetic field $H$ acting 
on the defect spins and directed along the coil axis that is taken along the 
axis $x$. Thus, the total Hamiltonian is
\be
\label{2}
 \hat H = \hat H_{def} - \mu_0 \sum_{j=1}^N \bB \cdot \bS_j \;  ,
\ee
in which $\mu_0$ is a defect magnetic moment and the total magnetic field
\be
\label{3}
\bB = B_0\bn_z + H \bn_x  
\ee
is the sum of a constant external field $B_0$ along the unit vector $\bn_z$
and of a coil magnetic field $H$ along the unit vector $\bn_x$. The magnetic 
moment $\mu_0$ is due to electrons and is negative. 

The coil magnetic field is a feedback field induced by the moving spins of 
the sample. The equation for this field follows from the Kirchhoff equation
and can be written \cite{Yukalov_41,Yukalov_42} in the form
\be
\label{4}
\frac{dH}{dt} + 2\gm H + \om^2 \int_0^t H(t')\; dt' = 
- 4\pi\; \frac{dm_x}{dt} \; ,
\ee
in which 
\be
\label{5}
\gm \equiv \frac{\om}{2Q}
\ee
is the circuit damping, $\omega$ is the circuit natural frequency, and $Q$ 
is a resonator quality factor. The effective electromotive force in the 
right-hand side of the equation is caused by the moving magnetization
\be
\label{6}
 m_x = \frac{\mu_0}{V_{res} } \sum_{j=1}^N \; \lgl S_j^x \rgl \;  ,
\ee
due to spins of the sample inside the coil of volume $V_{res}$. The angle
brackets imply statistical averaging.      

The equations of motion for spins are given by the Heisenberg equations
\be
\label{7}
  i\hbar \; \frac{dS_j^\nu}{dt} = 
\left [ S_j^\nu , \; \hat H \right ] \qquad ( \nu = x,y,z ) \;  .
\ee
Writing down these equations, we are interested in the temporal behavior
of the averaged over the sample spin polarizations
\be
\label{8}
 e_\nu \equiv \frac{1}{NS} \sum_{j=1}^N S_j^\nu  \qquad ( \nu = x,y,z ) \;  .
\ee
The latter are treated in the mean-field approximation with respect to 
statistical averaging:
\be
\label{9}
 \lgl e_\mu e_\nu \rgl = \lgl e_\mu  \rgl \lgl  e_\nu \rgl \; .
\ee
For the sake of simplicity, we write in what follows $e_\nu$ instead of
$<e_\nu>$. 

The feedback equation (4) defines the feedback magnetic field $H$, for 
which we introduce the dimensionless quantity
\be
\label{10}
 h \equiv \frac{H}{B_0}  \;  .
\ee
The electric circuit is tuned in resonance with the Zeeman frequency
\be
\label{11}
\om_0 \equiv \frac{1}{\hbar} \; | \mu_0 B_0 | \;  ,
\ee
so that 
\be
\label{12}
 \om = \om_0 \;  .
\ee
 
The parameter characterizing the coupling of the sample with the resonator 
can be defined as
\be
\label{13}
\bt \equiv \left | \; \frac{\mu_0 NS}{B_0 V_{res}} \; \right | \;   .
\ee
Differentiating Eq. (4), we come to the feedback equation
\be
\label{14}
\frac{d^2h}{dt^2} + \frac{1}{Q} \; \frac{dh}{dt} + h =
4\pi \bt  \; \frac{d^2 e_x}{dt^2} \;   ,
\ee
where time is measured in units of $1/ \omega$. The initial conditions
read as
\be
\label{15}
 h(0) = \dot{h}(0) = 0 \;  ,
\ee
with the overdot being time derivative. 

The total intensity of radiation can be estimated by the formula
$$
 I = \frac{2}{3c^3} \; \left | \ddot{\bf M} \right |^2 \;  ,
$$
where ${\bf M}$ is the sample magnetization moment
$$
{\bf M} = \mu_0 \sum_{j=1}^N \lgl \bf S_j \rgl \;   .
$$
As we have checked, the sharp peaks in the radiation intensity are due to coherent 
radiation, but when radiation is spread over a long time interval, it is mainly due 
to the incoherent part of radiation.  

In the following section, we present the results of numerical solution for
the equations describing the average spin polarizations $e_\nu$ and the corresponding 
radiation intensity. Our aim is to analyze the influence of the system parameters 
on the possibility of regulating the dynamics of graphene magnetization, paying 
the main attention to the influence of these parameters on the time of magnetization 
reversal, the value of the reversed spin polarization, and the related coherent spin
radiation.

\section{Defects at zigzag edge}

We consider the often met situation, when defects with spins are located 
along a zigzag edge of a graphene ribbon or a graphene flake. This edge is 
assumed to be directed along the axis $x$ corresponding to the direction
of the resonator coil. Such a geometry is taken because it guarantees the
best coupling of the resonator with the spins of the sample 
\cite{Yukalov_43,Yukalov_44}. The spins of the sample are prepared 
in a strongly nonequilibrium state, similarly to the setup employed in other
magnetic materials 
\cite{Yukalov_41,Yukalov_42,Yukalov_43,Yukalov_44,Davis_45,Henner_50}.
At the initial time, the defect spins are polarized downwards, with the total
initial polarization $e_z(0) = - 0.9$. This initial state is shown in Fig. 1. 
The sample is placed inside a resonator coil. And an external magnetic field 
is imposed, for which the equilibrium polarization would correspond to 
upwards spins. The questions of interest are: how quickly the spins reverse 
to the upward direction, how this reversal time depends on the system 
parameters, how the latter influence the value of the reversed magnetization,
as well as the strength of the radiation intensity.  

We accomplish numerical solution of the evolution equations for a chain of
$N = 100$ defect spins at a zigzag edge. The exchange interactions act only 
among the nearest neighbors, with the fixed strength $JS = \hbar \omega$, 
while other parameters are varied. In Fig. 2, we illustrate the role of 
magnetic anisotropy, characterized by the magnetic anisotropy parameter 
$\alpha$, for the coupling parameter $\beta = 0.01$ and the resonator quality 
factor $Q = 10$. Time is measured in units of $1/\omega$. Increasing the 
anisotropy makes the reversed value of $e_z$ smaller. Thus, when there is 
no anisotropy $(\alpha = 1)$, the polarization reversal is practically 
complete, with the final polarization $e_z \approx 0.9$. Under the anisotropy 
parameter $\alpha = 1.2$, the final $e_z \approx 0.75$, and for $\alpha = 1.4$, 
the final $e_z \approx 0.7$. The reversal time increases with the increase 
of the anisotropy. When there is no anisotropy $(\alpha = 1)$ the reversal 
time is $t_{rev} \approx 50$, although there are oscillations before the 
reversed magnetization stabilizes. For $\alpha = 1.2$, the reversal time 
is $t_{rev} \approx 100$. And for $\alpha = 1.4$, this time is 
$t_{rev} \approx 3000$. The anisotropy suppresses the radiation intensity,
as is clear form Fig. 3. In this and in the following figures, the radiation 
intensity is measured in units of erg/s = $10^{-7}$ W.  

The role of anisotropy can be diminished by a stronger coupling with the 
resonator, as is shown in Fig. 4, where $\beta = 0.1$ and $Q = 10$. In this 
figure, the first reversal happens at $t_{rev} \approx 15$ for all anisotropies 
up to $\alpha = 1.5$, although oscillations of $e_z$ remain for some time. The 
final reversed value of $e_z$ decreases with increasing anisotropy. The related 
radiation intensity is shown in Fig. 5. 

Increasing further the coupling between the sample and the resonator coil, 
although slightly diminishes the first reversal time, but induces strong 
oscillations of the spin polarizations, as is illustrated in Fig. 6 for 
$\beta = 1$ and $Q = 10$. The stronger coupling with the resonator increases
the radiation intensity, as is seen in Fig. 7. 

An intermediate situation occurs when the coupling is weak, but the quality 
factor is large, say, $\beta = 0.01$ and $Q = 100$. Then, at small anisotropy, 
there exist strong oscillations that become suppressed at increasing 
anisotropy. However, increasing the anisotropy essentially delays the 
magnetization reversal, as is shown in Fig. 8, and destroys the peak of the
radiation intensity, spreading it over a wide time interval, as follows from
Fig. 9.    

When for applications, it is necessary to achieve a short reversal time, but 
avoiding strong oscillations, then the optimal regime for this would correspond 
to not too strong coupling, under weak anisotropy, as in Figs. 2a and 2b. 
Although increasing the coupling suppresses the role of anisotropy, but leads 
to strong oscillations that could be undesirable in practical applications, 
such as information processing.    

However, when, on the contrary, it is necessary to increase the reversal 
time, which could be desirable for information preservation, this can be 
achieved by increasing the anisotropy. 

Since there are numerous ways of producing magnetic graphene by incorporating
different defects, as is discussed in the Introduction, the parameters of the
system can be varied in a wide range. Therefore this kind of magnetic graphene 
could provide possibility for different applications in spintronics, e.g., for 
information processing.

\section*{Acknowledgements}

Financial support from RFBR (grant 13-02-96018) and from Perm Ministry of 
Education (grant C-26/628) is appreciated. We are grateful for discussions 
to G. Sumanasekera and E.P. Yukalova.

\newpage

\begin{center}

{\Large {\bf Figure Captions}}

\end{center}

\vskip 2cm

{\bf Figure 1}. (a) Location of defects at a zigzag edge of a graphene 
ribbon: (b) initial configuration of defect spins. 

\vskip 5mm
{\bf Figure 2}. Transverse, $e_x$ (dotted line), and longitudinal, $e_z$ 
(solid line), spin polarizations as functions of dimensionless time 
(measured in units of $1 / \omega$), for the coupling parameter 
$\beta = 0.01$ and the resonator quality factor $Q = 10$, for different 
anisotropy parameters: (a) $\alpha = 1$, which implies the absence of 
anisotropy; (b) $\alpha = 1.2$; and (c) $\alpha = 1.4$. Increasing the 
anisotropy reduces the value of the reversed polarization $e_z$ and 
increases the reversal time.  

\vskip 5mm
{\bf Figure 3}. Radiation intensity in units erg/s = $10^{-7}$ W for the 
same sample parameters as in Fig. 2: (a) $\alpha = 1$ (absence of anisotropy); 
(b) $\alpha = 1.2$; and (c) $\alpha = 1.4$.  Increasing the anisotropy 
suppresses the radiation intensity. 

\vskip 5mm
{\bf Figure 4}. Spin polarizations $e_x$ (dotted line) and $e_z$ (solid line)
versus dimensionless time for the coupling parameter $\beta = 0.1$ and
the resonator quality factor $Q = 10$, for different anisotropy parameters:
(a) $\alpha = 1$ (no anisotropy); (b) $\alpha = 1.2$; and (c) $\alpha = 1.4$.
Increasing the anisotropy reduces the value of the reversed polarization 
$e_z$, but practically does not change the reversal time.  

\vskip 5mm
{\bf Figure 5}. Radiation intensity in units erg/s = $10^{-7}$ W for the 
same sample parameters as in Fig. 4: (a) $\alpha = 1$ (no anisotropy); 
(b) $\alpha = 1.2$; and (c) $\alpha = 1.4$. For a larger coupling parameter,
the suppression of radiation intensity by the anisotropy is weaker. 

\vskip 5mm
{\bf Figure 6}. Spin polarizations $e_x$ (dotted line) and $e_z$ (solid line)
versus dimensionless time for the coupling parameter $\beta = 1$ and the 
resonator quality factor $Q = 10$, for different anisotropy parameters:
(a) $\alpha = 1$ (no anisotropy); (b) $\alpha = 1.2$; and (c) $\alpha = 1.4$.
Increasing the anisotropy reduces the value of the reversed polarization 
$e_z$, almost does not change the reversal time, but induces strong 
oscillations of the spin polarizations.   

\vskip 5mm
{\bf Figure 7}. Radiation intensity in units erg/s = $10^{-7}$ W for the 
same sample parameters as in Fig. 6: (a) $\alpha = 1$ (no anisotropy); 
(b) $\alpha = 1.2$; and (c) $\alpha = 1.4$.  For a larger coupling parameter,
the radiation intensity is not strongly suppressed by the anisotropy. 

\vskip 5mm
{\bf Figure 8}. Spin polarizations $e_x$ (dotted line) and $e_z$ (solid line)
versus dimensionless time for the coupling parameter $\beta = 0.01$ and 
the resonator quality factor $Q = 100$, for different anisotropy parameters:
(a) $\alpha = 1$ (no anisotropy); (b) $\alpha = 1.2$; and (c) $\alpha = 1.4$.
Increasing the anisotropy suppresses spin oscillations, only slightly 
reduces the value of the reversed polarization $e_z$, but strongly delays 
the reversal time. 

\vskip 5mm
{\bf Figure 9}. Radiation intensity in units erg/s = $10^{-7}$ W for the 
same sample parameters as in Fig. 8: (a) $\alpha = 1$ (no anisotropy); 
(b) $\alpha = 1.2$; and (c) $\alpha = 1.4$. Increasing the resonator quality 
factor induces many oscillations in radiation intensity.

\newpage

\begin{figure}[ht]
\centerline{\includegraphics[width=12cm]{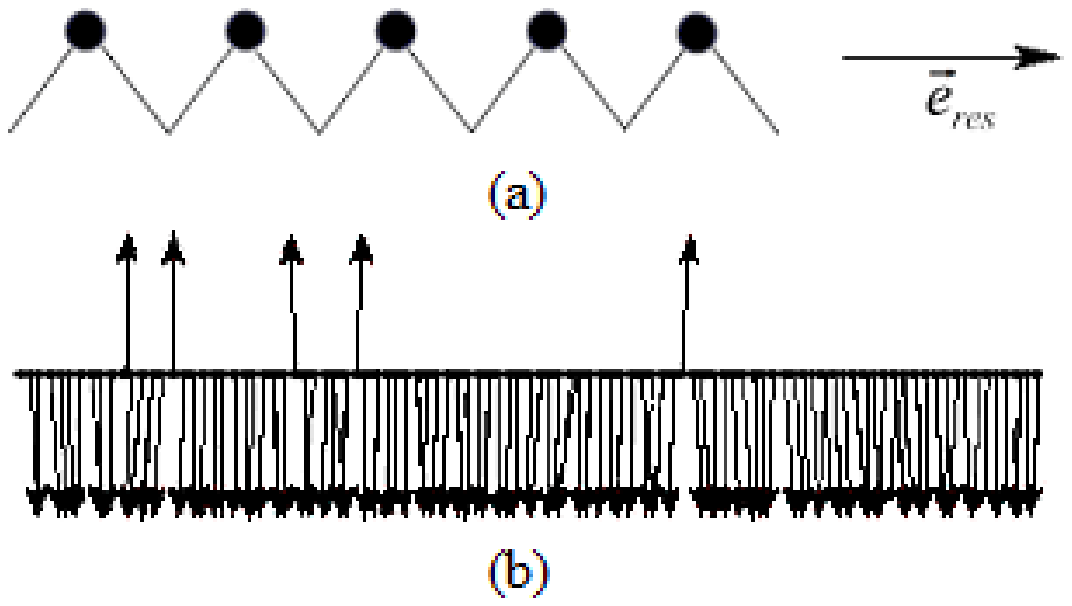} }
\caption{(a) Location of defects at a zigzag edge of a graphene 
ribbon: (b) initial configuration of defect spins.
}
\label{fig:Fig.1}
\end{figure}

\newpage

\begin{figure}[ht]
\centerline{\includegraphics[width=12cm]{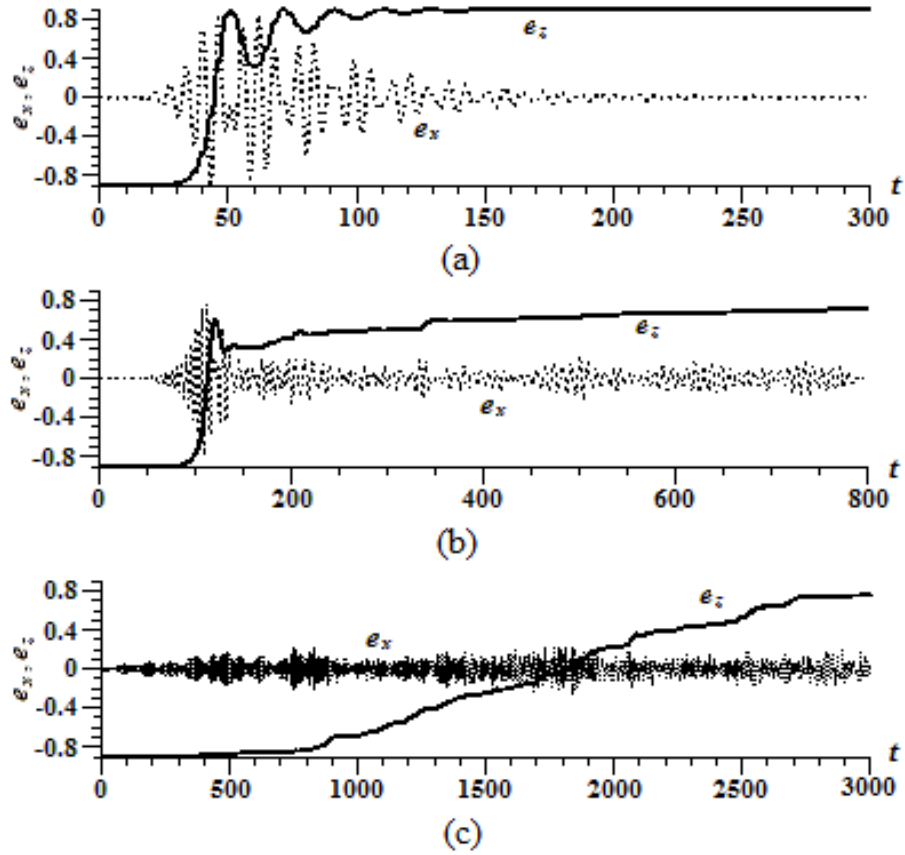} }
\caption{Transverse, $e_x$ (dotted line), and longitudinal, $e_z$ 
(solid line), spin polarizations as functions of dimensionless time 
(measured in units of $1 / \omega$), for the coupling parameter 
$\beta = 0.01$ and the resonator quality factor $Q = 10$, for different 
anisotropy parameters: (a) $\alpha = 1$, which implies the absence of 
anisotropy; (b) $\alpha = 1.2$; and (c) $\alpha = 1.4$. Increasing the 
anisotropy reduces the value of the reversed polarization $e_z$ and 
increases the reversal time. 
}
\label{fig:Fig.2}
\end{figure}

\newpage

\begin{figure}[ht]
\centerline{\includegraphics[width=12cm]{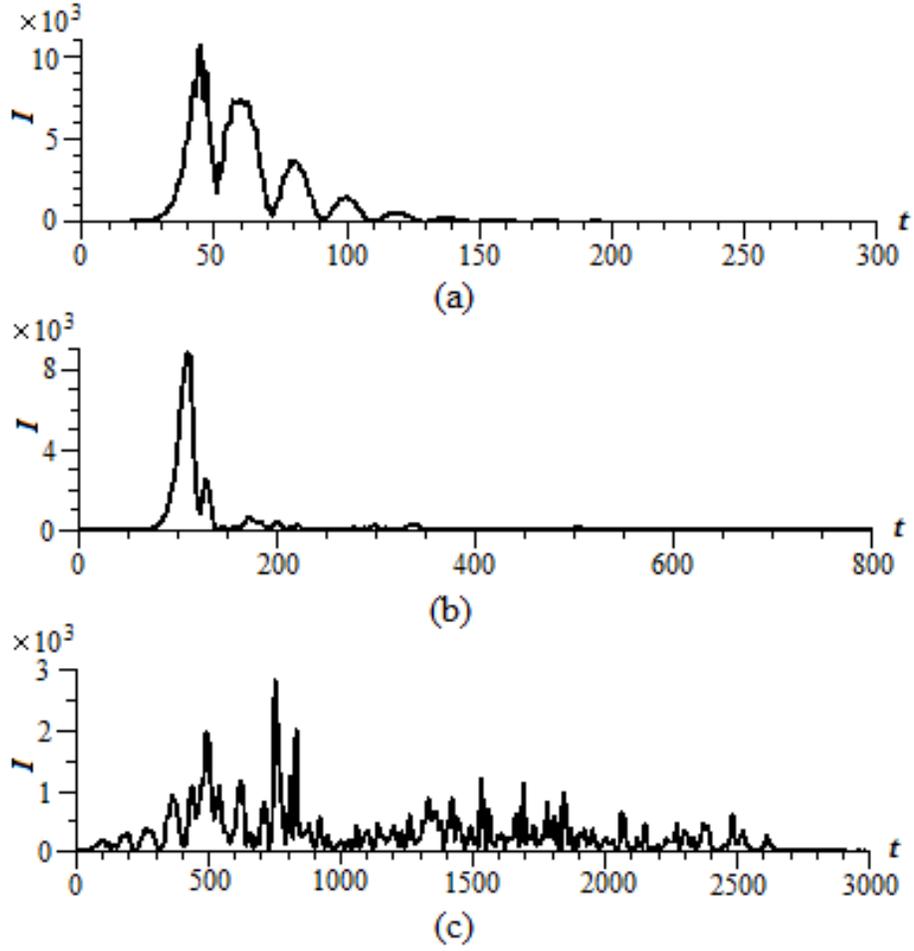} }
\caption{Radiation intensity in units erg/s = $10^{-7}$ W for the 
same sample parameters as in Fig. 2: (a) $\alpha = 1$ (absence of anisotropy); 
(b) $\alpha = 1.2$; and (c) $\alpha = 1.4$.  Increasing the anisotropy 
suppresses the radiation intensity.
}
\label{fig:Fig.3}
\end{figure}

\newpage

\begin{figure}[ht]
\centerline{\includegraphics[width=12cm]{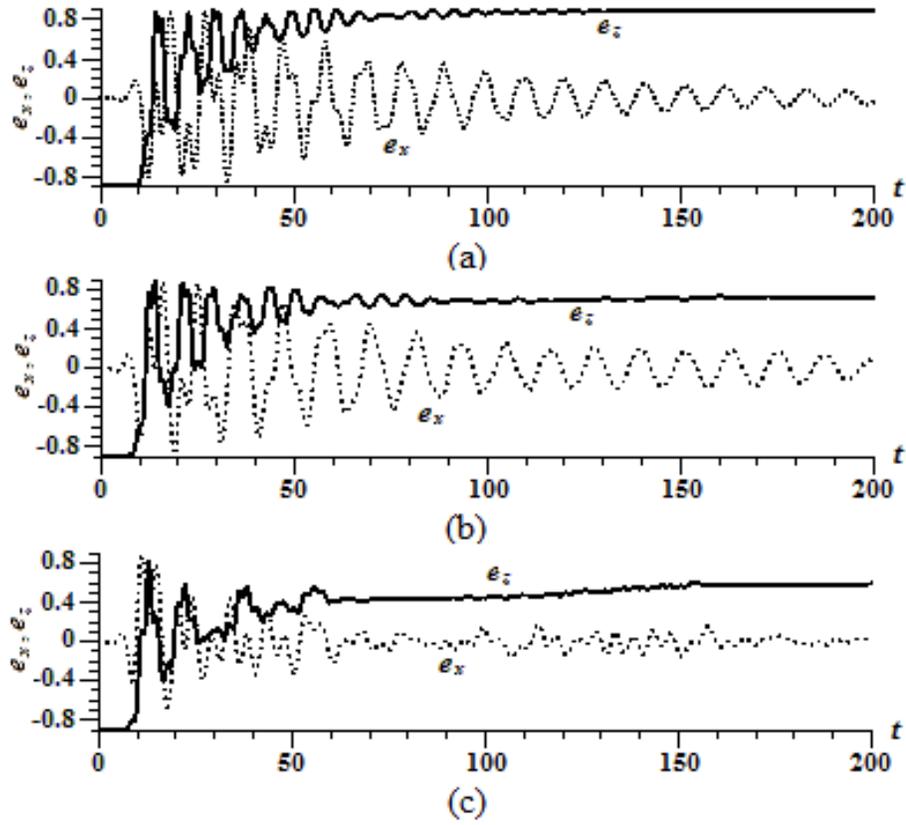} }
\caption{Spin polarizations $e_x$ (dotted line) and $e_z$ (solid line)
versus dimensionless time for the coupling parameter $\beta = 0.1$ and
the resonator quality factor $Q = 10$, for different anisotropy parameters:
(a) $\alpha = 1$ (no anisotropy); (b) $\alpha = 1.2$; and (c) $\alpha = 1.4$.
Increasing the anisotropy reduces the value of the reversed polarization 
$e_z$, but practically does not change the reversal time. 
}
\label{fig:Fig.4}
\end{figure}

\newpage

\begin{figure}[ht]
\centerline{\includegraphics[width=12cm]{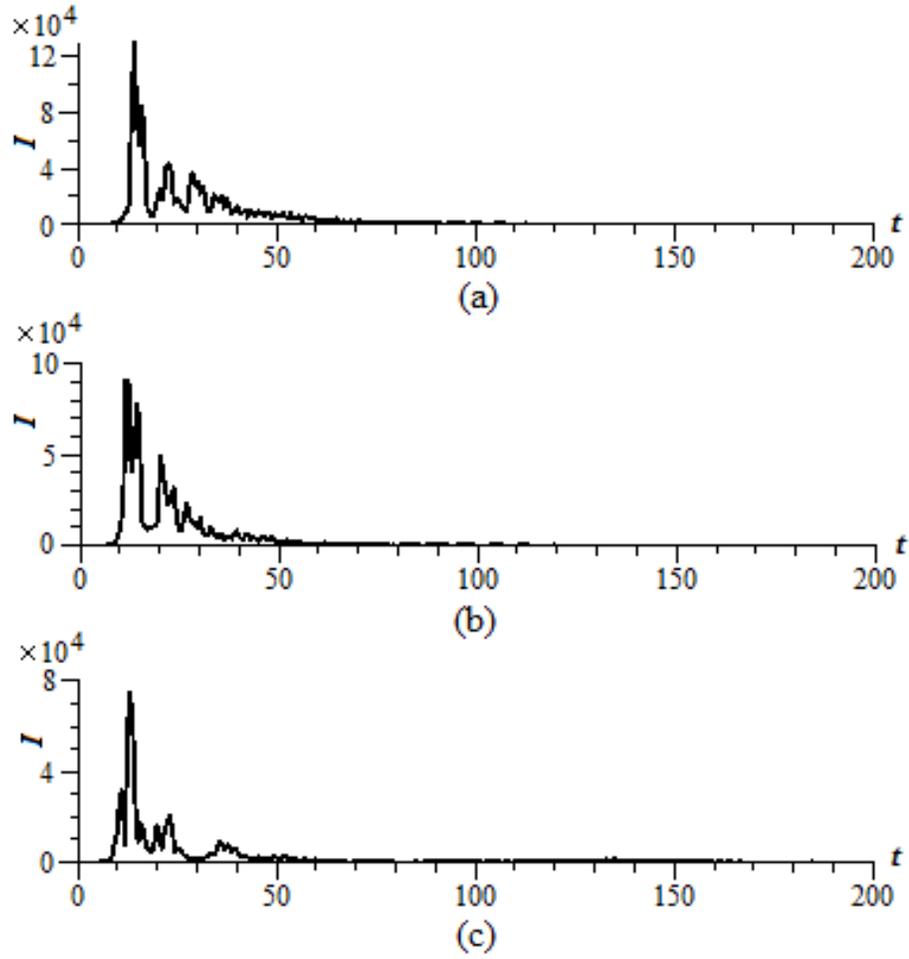} }
\caption{Radiation intensity in units erg/s = $10^{-7}$ W for the 
same sample parameters as in Fig. 4: (a) $\alpha = 1$ (no anisotropy); 
(b) $\alpha = 1.2$; and (c) $\alpha = 1.4$.  For a larger coupling parameter,
the suppression of radiation intensity by the anisotropy is weaker. 
}
\label{fig:Fig.5}
\end{figure}

\newpage

\begin{figure}[ht]
\centerline{\includegraphics[width=12cm]{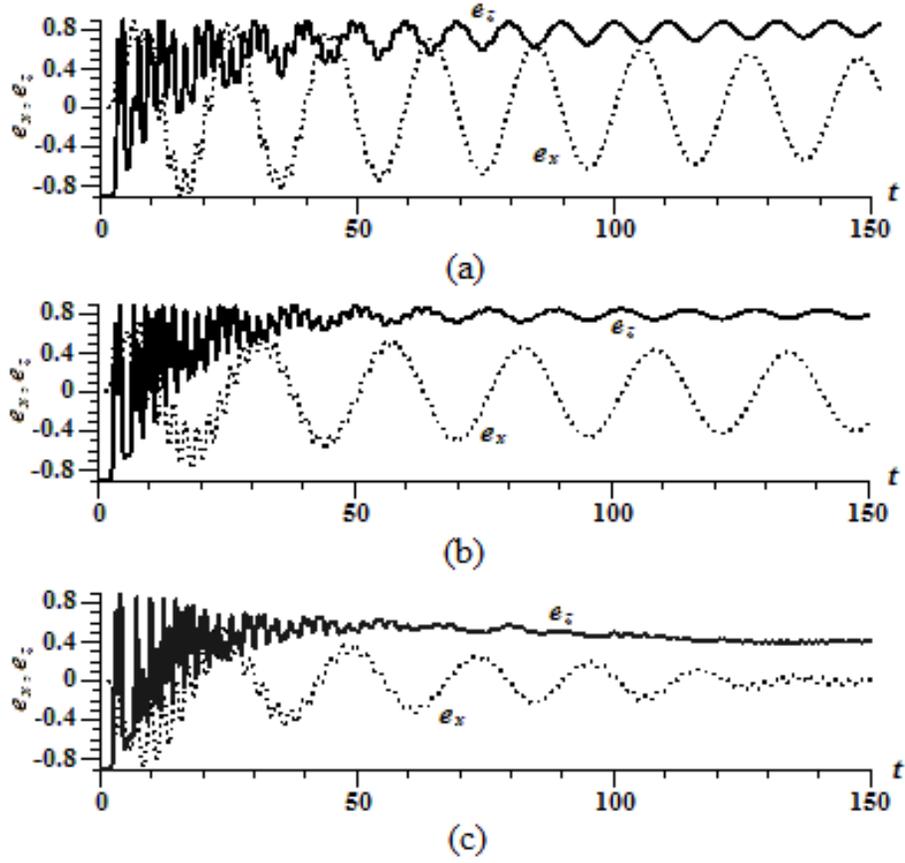} }
\caption{Spin polarizations $e_x$ (dotted line) and $e_z$ (solid line)
versus dimensionless time for the coupling parameter $\beta = 1$ and the 
resonator quality factor $Q = 10$, for different anisotropy parameters:
(a) $\alpha = 1$ (no anisotropy); (b) $\alpha = 1.2$; and (c) $\alpha = 1.4$.
Increasing the anisotropy reduces the value of the reversed polarization 
$e_z$, almost does not change the reversal time, but induces strong 
oscillations of the spin polarizations. 
}
\label{fig:Fig.6}
\end{figure}

\newpage

\begin{figure}[ht]
\centerline{\includegraphics[width=12cm]{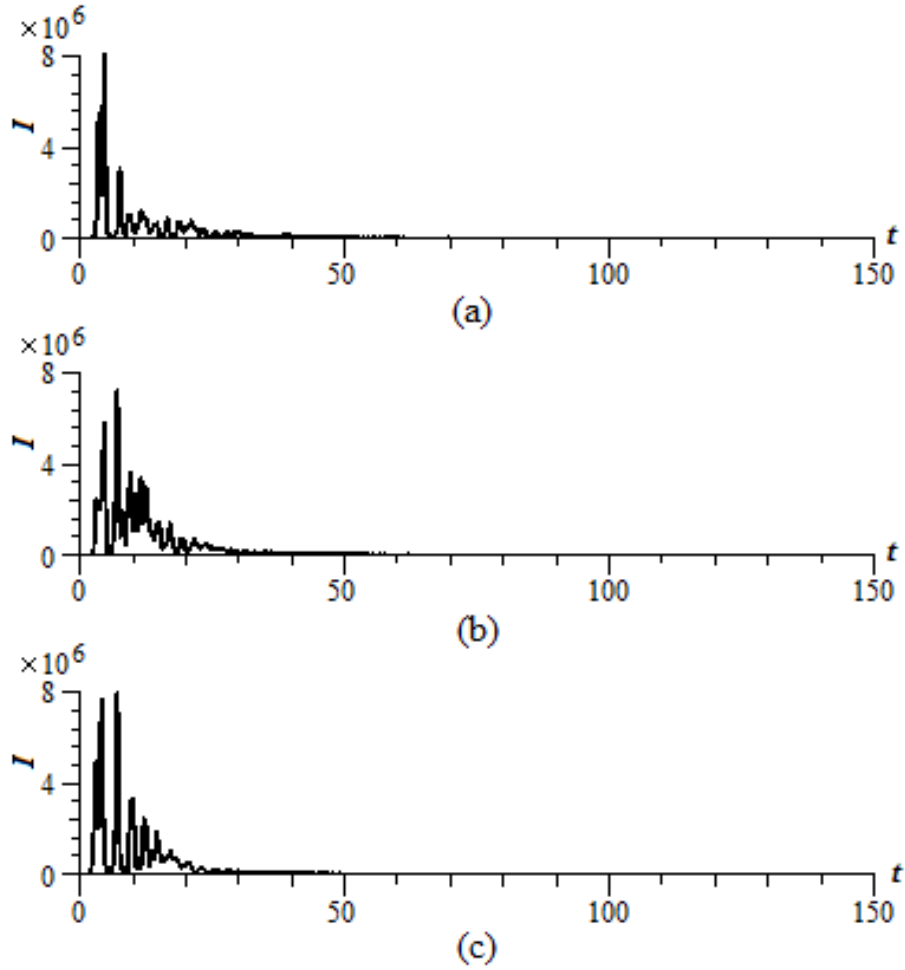} }
\caption{Radiation intensity in units erg/s = $10^{-7}$ W for the 
same sample parameters as in Fig. 6: (a) $\alpha = 1$ (no anisotropy); 
(b) $\alpha = 1.2$; and (c) $\alpha = 1.4$.  For a larger coupling parameter,
the radiation intensity is not strongly suppressed by the anisotropy. 
}
\label{fig:Fig.7}
\end{figure}

\newpage

\begin{figure}[ht]
\centerline{\includegraphics[width=12cm]{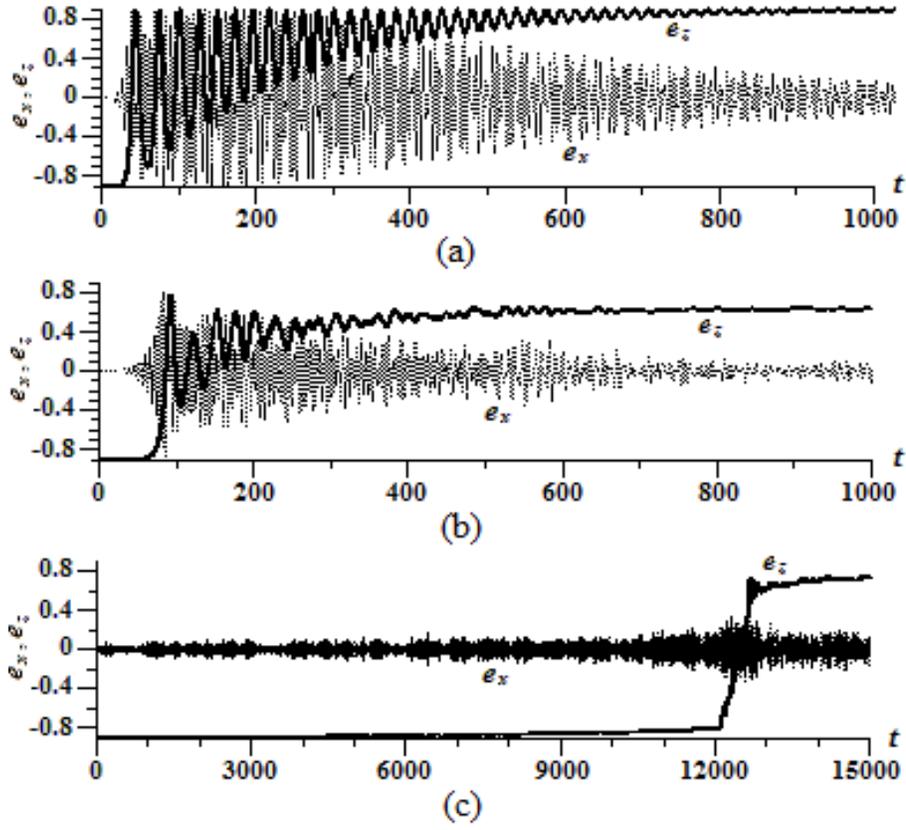} }
\caption{Spin polarizations $e_x$ (dotted line) and $e_z$ (solid line)
versus dimensionless time for the coupling parameter $\beta = 0.01$ and 
the resonator quality factor $Q = 100$, for different anisotropy parameters:
(a) $\alpha = 1$ (no anisotropy); (b) $\alpha = 1.2$; and (c) $\alpha = 1.4$.
Increasing the anisotropy suppresses spin oscillations, only slightly 
reduces the value of the reversed polarization $e_z$, but strongly delays 
the reversal time.
}
\label{fig:Fig.8}
\end{figure}

\newpage

\begin{figure}[ht]
\centerline{\includegraphics[width=12cm]{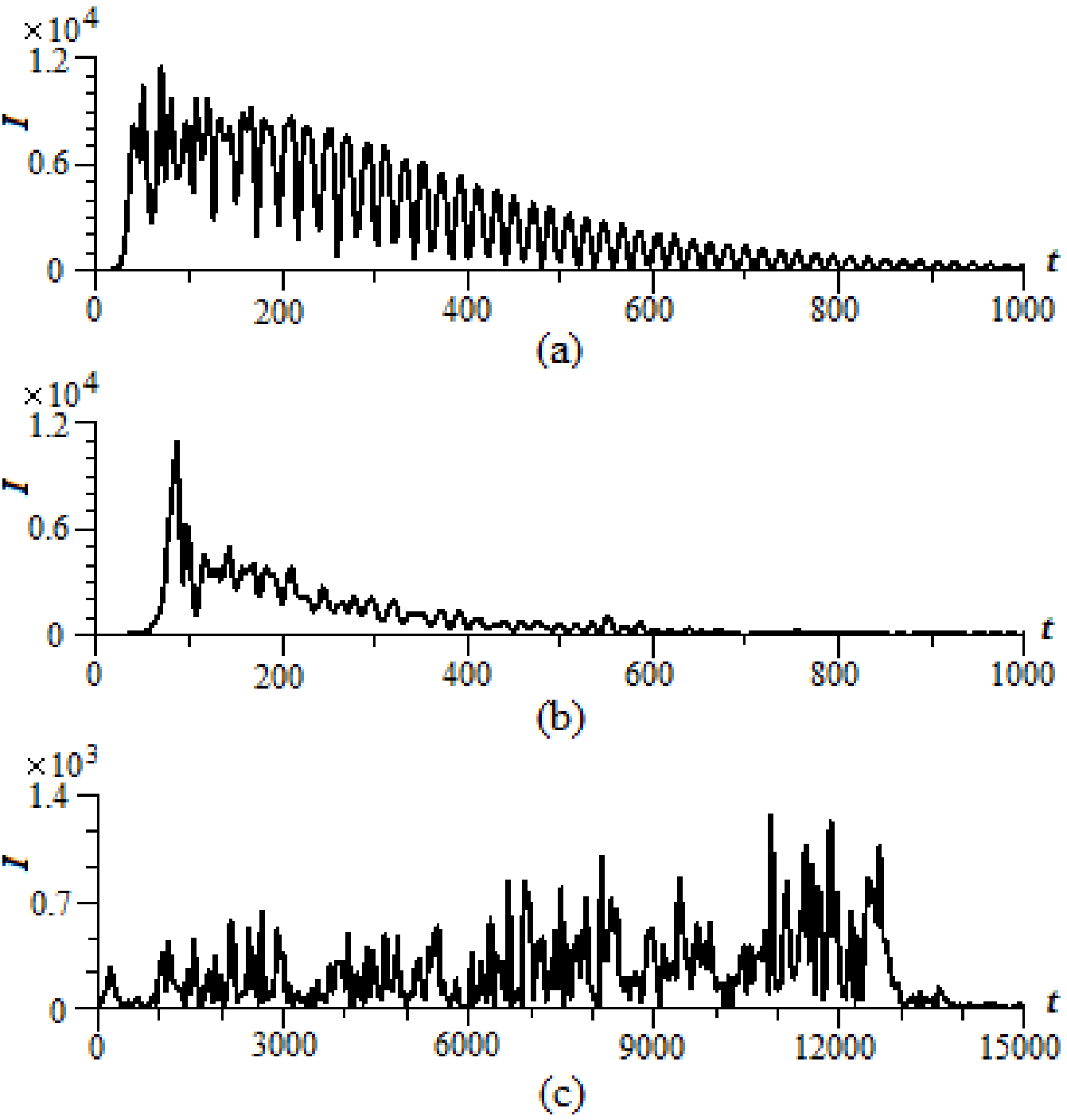} }
\caption{Radiation intensity in units erg/s = $10^{-7}$ W for the 
same sample parameters as in Fig. 8: (a) $\alpha = 1$ (no anisotropy); 
(b) $\alpha = 1.2$; and (c) $\alpha = 1.4$. Increasing the resonator quality 
factor induces many oscillations in radiation intensity. 
}
\label{fig:Fig.9}
\end{figure}


\newpage


\begin{thebibliography}{99}

\bibitem{Birman_1}
Birman J L, Nazmitdinov R G and Yukalov V I  2013  
{\it Phys. Rep.} {\bf 526} 1 

\bibitem{Castro_1}
Castro Neto A H, Guinea F, Peres N M, Novoselov K S and Geim A K 2009
{\it Rev. Mod. Phys.} {\bf 81} 109 

\bibitem{Abergel_2}
Abergel D S, Apalkov V, Berashevich J, Ziegler K and Chakraborty T  2010
{\it Adv. Phys.} {\bf 59} 261 

\bibitem{Dresselhaus_38}
Dresselhaus M S, Jorio A and Saito R  2010
{\it Annu. Rev. Condens. Matter Phys.} {\bf 1} 89  

\bibitem{Goerbig_3}
Goerbig M O  2011 
{\it Rev. Mod. Phys.} {\bf 83} 1193 

\bibitem{Saito_39}
Saito R, Hofman M, Dresselhaus G, Jorio A and Dresselhaus M S  2011
{\it Adv. Phys.} {\bf 60} 413 

\bibitem{Kotov_4}
Kotov V N, Uchoa B, Pereira V M, Guinea F and Castro Neto A H  2012
{\it Rev. Mod. Phys.} {\bf 84} 1067  

\bibitem{Katsnelson_5}
Katsnelson M I  2012 
{\it Graphene: Carbon in Two Dimensions} (Cambridge: Cambridge University) 

\bibitem{Mccann_6}
McCann E and Koshiro M  2013
{\it Rep. Prog. Phys.} {\bf 76} 056503 

\bibitem{Wehling_7}
Wehling T O, Black-Schaffer A M and Balatsky A V  2014
{\it Adv. Phys.} {\bf 63} 1  

\bibitem{Palacios_46}
Palacios J J  2010
{\it Semicond. Sci. Technol.} {\bf 25} 033003 

\bibitem{Andrei_8}
Andrei N, Furuya K and Lowenstein J H  1983
{\it Rev. Mod. Phys.} {\bf 55} 331  

\bibitem{Huang_9}
Huang L, Lai Y C, Ferry D K, Akis R and Goodnick S M  2009
{\it J. Phys. Condens. Matter} {\bf 21} 344203 

\bibitem{Enoki_10}
Enoki T, Kobayashi Y  2005
{\it J. Mater. Chem.} {\bf 15} 3999

\bibitem{Jiang_11}
Jiang D E, Sumpter B G and Dai S  2007
{\it J. Chem. Phys.} {\bf 127} 124703 

\bibitem{Otani_12}
Otani M, Koshino M, Takagi Y and Okada S  2010
{\it Phys. Rev. B} {\bf 81} 161403   

\bibitem{Gorjizadeh_13}
Gorjizadeh N, Ota N and Kawazoe Y  2013
{\it Chem. Phys.} {\bf 415} 64 

\bibitem{Magda_14}
Magda G Z, Jin X, Hagymasi I, Vancso P, Osvath Z, Nemes-Incze P, Hwang C, 
Biro L P and Tapaszto L  2014 
{\it Nature} {\bf 514} 608

\bibitem{Lisenkov_15}
Lisenkov S, Andriotis A N and Menon M  2010
{\it Phys. Rev. B} {\bf 82} 165454  

\bibitem{Chen_16}
Chen J H, Li L, Cullen W G, Williams E D and Fuhrer M S  2011
{\it Nature Phys.} {\bf 7} 535  

\bibitem{Nair_47}
Nair R R, Sepioni M, Tsai I L, Lehtinen O, Keinonen J, Krasheninnikov A V, Thomson T,
Geim A K, and Grigorieva I V  2012
{\it Nature Phys.} {\bf 8} 199 

\bibitem{Yang_18}
Yang H X, Chshiev M, Boukhvalov D W, Waintal X and Roche S  2011
{\it Phys. Rev. B} {\bf 84} 214404  

\bibitem{Lopez_19}
Lopez-Sancho M P, Castro E V and Vozmediano M A  2013
{\it AIP Conf. Proc.} {\bf 1517} 229 

\bibitem{Li_20}
Li Y, He J and Kou S P  2014
{\it Phys. Rev. B}  {\bf 90} 201406

\bibitem{Faccio_21}
Faccio R and Mombru A W  2015
{\it Comput. Mater. Sci.} {\bf 97} 193 

\bibitem{Bhandary_48}
Bhandary S, Eriksson O and Sanyal B  2013
{\it Sci. Rep.} {\bf 3} 3405 

\bibitem{Jaskolski_40}
Jaskolski W, Chio L and Ayuela A  2015
{\it arXiv 1503.08750}  

\bibitem{Power_22}
Power S R, de Menezes V M, Fagan S B and Fereira M S  2011
{\it Phys. Rev. B} {\bf 84} 195421 

\bibitem{Pike_23}
Pike N A and Stroud D  2014
{\it Phys. Rev. B} {\bf 89} 115428 

\bibitem{Pike_24}
Pike N A and Stroud D  2014
{\it Appl. Phys. Lett.} {\bf 105} 052404  

\bibitem{Blonsky_49}
Blonsky P and Hafner J  2014
{\it J. Phys. Condens. Matter} {\bf 26} 256001 

\bibitem{Kabir_25}
Kabir M and Saha-Dasgupta T  2014
{\it Phys. Rev. B} {\bf 90} 064420

\bibitem{Qaiumzadeh_26}
Qaiumzadeh A and Asgari R  2009
{\it Phys. Rev. B} {\bf 80} 035429 

\bibitem{Zhang_27}
Zhang Z, Chen C, Zeng X C and Guo W  2010 
{\it Phys. Rev. B} {\bf 81} 155428

\bibitem{Garnica_28}
Garnica M, Stradi D, Barja S, Calleja F, Dias C, Alcami M, Martin N,
Vazquez de Parga A L, Martin F and Miranda R  2013
{\it Nature Phys.} {\bf 9} 368 

\bibitem{Hu_29}
Hu T, Zhou J, Dong J M and Kawazoe Y  2013
{\it Phys. Rev. B} {\bf 87} 079906 

\bibitem{Roy_30}
Roy B, Assaad F F and Herbut I F  2014
{\it Phys. Rev. X} {\bf 4} 021042

\bibitem{Boukhvalov_31}
Boukhvalov D W and Katsnelson M I  2011
{\it ACS Nano} {\bf 5} 2440

\bibitem{Panich_32}
Panich A M, Shames A I and Sergeev N A  2013
{\it Appl. Magn. Res.} {\bf 44} 107   

\bibitem{Terrones_33}
Terrones H, Lv R, Terrones M and Dresselhaus M S  2012
{\it Rep. Prog. Phys.} {\bf 75} 062501  

\bibitem{Bekyarova_34}
Bekyarova E, Sarkar S, Niyogi S, Itkis M E and Haddon R C  2012
{\it J. Phys. D: Appl. Phys.} {\bf 45} 154009  

\bibitem{Yaziev_35}
Yaziev  O V  2010
{\it Rep. Prog. Phys.} {\it 73} 056501 

\bibitem{Enoki_36}
Enoki T and Ando T  2013
{\it Physics and Chemistry of Graphene} (Singapore: Pan Stanford) 

\bibitem{Purcell_37}
Purcell E M  1946
{\it Phys. Rev.} {\bf 69} 681 

\bibitem{Yukalov_41}
Yukalov V I  1996
{\it Phys. Rev. B} {\bf 53} 9232

\bibitem{Yukalov_42}
Yukalov V I and Yukalova E P  2004
{\it Phys. Part. Nucl.} {\bf 35} 348 

\bibitem{Yukalov_43}
Yukalov V I, Henner V K, Kharebov P V and Yukalova E P  2008
{\it Laser Phys. Lett.} {\bf 5} 887

\bibitem{Yukalov_44}
Yukalov V I, Henner V K and Yukalova E P  2015
{\it J. Phys. Conf. Ser.} {\bf 594} 012006 

\bibitem{Davis_45}
Davis C L, Kaganov I V and Henner V K  2000 
{\it Phys. Rev. B} {\bf 62} 12328

\bibitem{Henner_50}
Henner V, Raikher Y and Kharebov P  2011
{\it Phys. Rev. B} {\bf 84} 144412


\end{thebibliography}
\end{document}